\documentclass[a4paper]{article}
\usepackage{amsmath,graphicx}
\usepackage{amsfonts}
\usepackage{amssymb}
\usepackage{geometry}

\usepackage{caption}
\usepackage{subcaption}

\date{}

\title{DEEP JOINT SOURCE-CHANNEL CODING FOR WIRELESS IMAGE RETRIEVAL}

\author{Mikolaj Jankowski, Deniz G{\"u}nd{\"u}z, Krystian Mikolajczyk\\
Imperial College London\\
{\tt\small \{mikolaj.jankowski17, d.gunduz, k.mikolajczyk\}@imperial.ac.uk}}

\begin{document}

\maketitle

\begin{abstract}
Motivated by surveillance applications with wireless cameras or drones, we consider the problem of image retrieval over a wireless channel. Conventional systems apply lossy compression on query images to reduce the data that must be transmitted over the bandwidth and power limited wireless link. We first note that reconstructing the original image is not needed for retrieval tasks; hence, we introduce a deep neutral network (DNN) based compression scheme targeting the retrieval task. Then, we completely remove the compression step, and propose another DNN-based communication scheme that directly maps the feature vectors to channel inputs. This joint source-channel coding (JSCC) approach not only improves the end-to-end accuracy, but also simplifies and speeds up the encoding operation which is highly beneficial for power and latency constrained IoT applications.
\end{abstract}


\section{Introduction}
\label{sec:introduction}

Internet of Things (IoT) devices have become  widespread in recent years. Typically, these are small, non-standard computers designed to perform certain tasks, including measurement, recording, or computing, and use wireless links to transmit their measurements. Since the transmission power is the main source of energy consumption for IoT devices, they typically employ  compression methods to reduce the amount of data they transmit, while retaining the information required to achieve the underlying goal. There are many lossless/ lossy compression techniques in the literature for various types of information sources, e.g. audio, image, video, etc. However, in many IoT applications, the receiver does not require the entire source data, as its goal is typically to use some of the information carried by the data. Such application scenarios can benefit from novel and more effective task-oriented compression schemes \cite{resonance, torfason2018towards}.   

In this paper, we consider one of the most challenging retrieval tasks -- person re-identification (re-ID), carried out over a wireless channel. It aims at matching a query image of a person recorded by a remote wireless camera, to an image of the same person, stored in a large pedestrian database (gallery) available at the access point. The matching is typically done by  extracting features from images and then computing the similarity score between the features rather than the source images. We consider two types of approaches. In the  ``digital'' scheme, the features are first compressed for the task, and the compressed bits are encoded with a channel code for reliable transmission. This approach suffers from the \textit{cliff effect}: desired performance achieved only at the target channel quality, and degrades sharply if the experienced channel is worse than the target value. Therefore, the performance of digital schemes depend critically on accurate channel estimation, and requires high performance codes for channel coding and compression. Such codes are available for long blocklengths, hence not appropriate for some of the low-latency IoT applications. 

The second approach is based on joint source-channel coding (JSCC), which does not require converting either the image or the feature vector into bits. Instead, the source measurements are directly mapped to channel symbols. A deep neural network (DNN) based JSCC has recently been shown to outperform state-of-the-art digital schemes for wireless image transmission \cite{jscc_image_dnn}. Here, we consider both digital and JSCC architectures for image retrieval over wireless channels, which can be considered as task-based JSCC. Our contributions can be summarized as follows:
\begin{itemize}
    \item We propose a task-based compression scheme for input images for the re-ID task, which combines a re-ID baseline with a feature encoder, followed by scalar quantization and entropy coding.
    \item We propose an autoencoder-based architecture and training strategy for robust JSCC of feature vectors, generated by a retrieval baseline, under noisy and bandwidth-limited channel conditions.
    \item We perform extensive evaluations under different signal-to-noise ratio (SNR) and bandwidth constraints, and show that the proposed JSCC scheme outperforms digital scheme.
    \item We evaluate the proposed schemes on the person re-ID task, and show that the performance close to the noiseless bound can be achieved even under very harsh SNR and bandwidth constraints.
\end{itemize}

 Recently, JSCC was evaluated for a classification task in \cite{jscc_features}. However, in classification it is not necessary to send feature vectors, as the transmitter can perform the task locally and send only the class label. This requires the reliable transmission of only $\log_2 10$ bits for the CIFAR10 dataset considered in \cite{jscc_features}. In contrast, in the retrieval problem, the transmitter does not have access to the gallery, thus cannot compute the similarity scores between the query and the gallery images locally.

\section{Methods}

In this work we consider both the digital (separate) and the JSCC approaches. In both methods, feature vectors are first extracted from the images as low-dimensional representation of human identities (Section \ref{subs:prid_baseline}), and are transmitted over the wireless channel. The features cannot be transmitted in a lossless fashion due to finite channel capacity. The recovered feature vector at the receiver is compared to the vectors in a local image database, called the \textit{gallery}, to find the nearest neighbour, thus to re-ID the person. Note that we do not consider the traditional image compression schemes, i.e., transmitting the images directly instead of features, as our bandwidth limitations prevent from sending even highly compressed images.

While our proposed JSCC method can be trained with any differentiable channel model, we consider an additive white Gaussian noise (AWGN) channel in this paper. In particular, given a channel input vector $\mathbf{x} \in \mathbb{R}^B$, the channel output vector $\mathbf{y} \in \mathbb{R}^B$ is given by $\mathbf{y} = \mathbf{x} + \mathbf{z}$, where $\mathbf{z}$ is the noise vector consisting of independent and identically distributed noise component, drawn from a zero-mean normal distribution with variance $\sigma^2$. We impose an average power constraint of $P$ for every channel input vector, i.e., $\frac{1}{B} \sum_{i=1}^B x_i^2 \leq P$. We evaluate the re-ID performance for different channel SNRs given by $\frac{P}{\sigma^2}$. To calculate the minimum $\text{SNR}$ requirement for the digital scheme, we use the Shannon capacity formula $C = \frac{1}{2}\log_2\left(1 + \frac{P}{\sigma^2}\right)$.

\subsection{Person Re-ID Baseline}
\label{subs:prid_baseline}

Following the state-of-the-art person re-ID methods \cite{pyramid, DBLP:journals/corr/abs-1711-09349, mgn, triplet} we employ the ResNet-50 network \cite{resnet}, pretrained on ImageNet \cite{imagenet}, for feature extraction. This ensures that similar results can be expected in different setups. In more detail, we use ResNet-50 with batch normalization layers applied after each convolution. As input, we use images resized to a common $256 \times 128$ resolution with bicubic interpolation. For the last layer we use average pooling across all the channels, which results in a 2048-dimensional feature vector. During training we use stochastic gradient descent (SGD) with a learning rate of $0.01$ and a momentum of $0.9$. We also apply $L_2$ regularization, weighted by $5\cdot 10^{-4}$ to the ResNet-50 parameters. We refer to this architecture as the \textit{re-ID baseline}.

\subsection{Digital Transmission of Compressed Feature Vectors}
\label{subs:approach3}

\begin{figure}[]
\begin{center}
\includegraphics[width=0.95\linewidth]{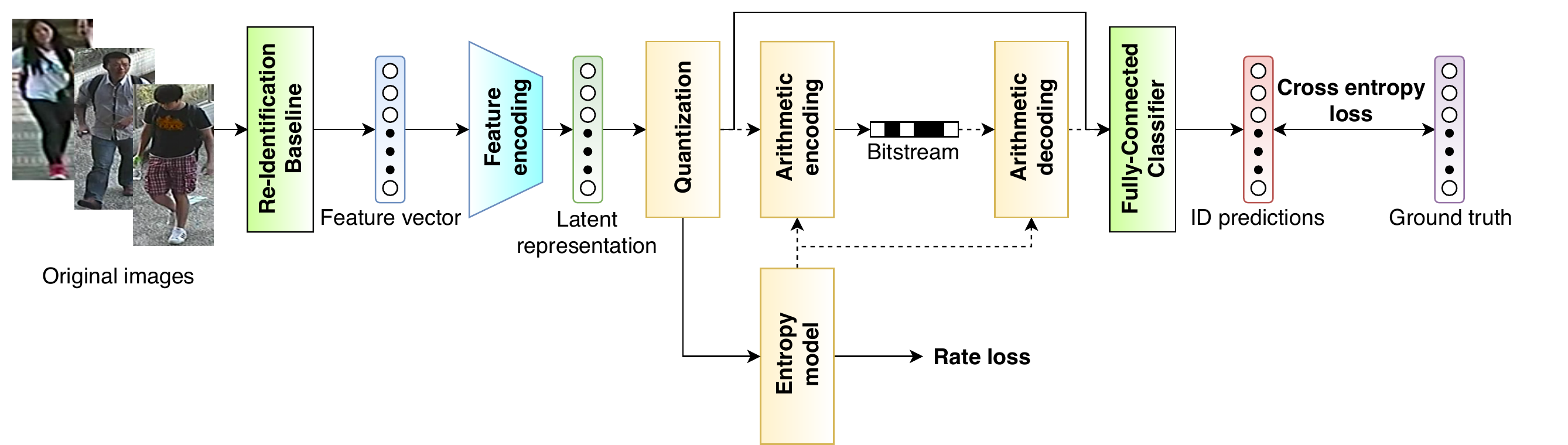}
\end{center}
  \caption{The digital transmission scheme. Input is transformed into a feature vector, which is compressed using a DNN. At the receiver, latent representation is classified into IDs to compute the loss. Arithmetic coding is bypassed during training. Channel coding step is not shown for better readability.}
\label{fig:approach3}
\end{figure}

An overview of the proposed digital scheme is shown in Fig. \ref{fig:approach3}.  We first extract features using the re-ID baseline described in Section \ref{subs:prid_baseline}. The feature vector is then compressed into as few bits as possible through lossy compression followed by arithmetic coding. The compressed bits are then channel coded, with introduced structured redundancy to counter the channel noise. 
The lossy feature encoder consists of a single fully-connected layer for dimensionality reduction, followed by quantization. On the receiver side we use the quantized latent representation as a feature vector, which is passed through a fully-connected layer for ID classification. Note that the IDs are used for calculating the loss during training only. During retrieval, the feature vectors are used for nearest neighbour search.

For quantization, we adopt the quantization noise from \cite{quantization} as a differentiable approximation of this operation during training. In order to model the density of the unknown prior $P_q$ of the low-dimensional quantized representation we adopt a flexible model based on its cumulative distribution function from \cite{balle2018variational}. The model consists of learnable parameters and can be trained together with a neural network to assign likelihood of occurrence of each element within the latent representation. Probability distribution estimated this way is subsequently used to compute Shannon entropy of the latent representation and produce rate loss which is jointly minimized with cross-entropy loss. To enable a smooth trade-off between re-ID performance and the compression rate,  we minimize the weighted loss of the two objectives:
\begin{equation}
    L = -\log_2P_q + \lambda \cdot l_{ce},
\end{equation}
where $l_{ce}$ is the cross-entropy between predicted classes (identities) and ground truth for the person re-ID task. The other part of the loss function corresponds to the Shannon entropy of the quantized vector.

For training the feature encoder, the fully-connected classifier and the density model, we use SGD with learning rate $0.01$ and momentum $0.9$. We further apply $L_2$ regularizer to the encoder parameters, weighted by $5\cdot 10^{-3}$. We train the whole network for 30 epochs, reduce the learning rate to $0.001$ and train for further 30 epochs.

The latents obtained after the quantization step are arithmetically encoded, and transmitted with a channel code. Note that any channel code will introduce some errors; therefore, there is an inherent trade-off between the compression rate and the channel coding rate for a given constraint on the channel bandwidth. Compressing the feature vector further leads to increased distortion, but  also allows to introduce more redundancy, and hence, increased reliability against noise. In general, the optimal compression and channel coding rates depend on the distortion-rate function of the compression scheme and the error-rate of the channel code. To simplify this task, we assume  capacity-achieving channel codes over the channel, as well as  reliable transmissions. This  provides an upper bound on the performance that can be achieved by any digital scheme that uses the above architecture. 

\begin{figure}[]
\begin{center}
\includegraphics[width=0.95\linewidth]{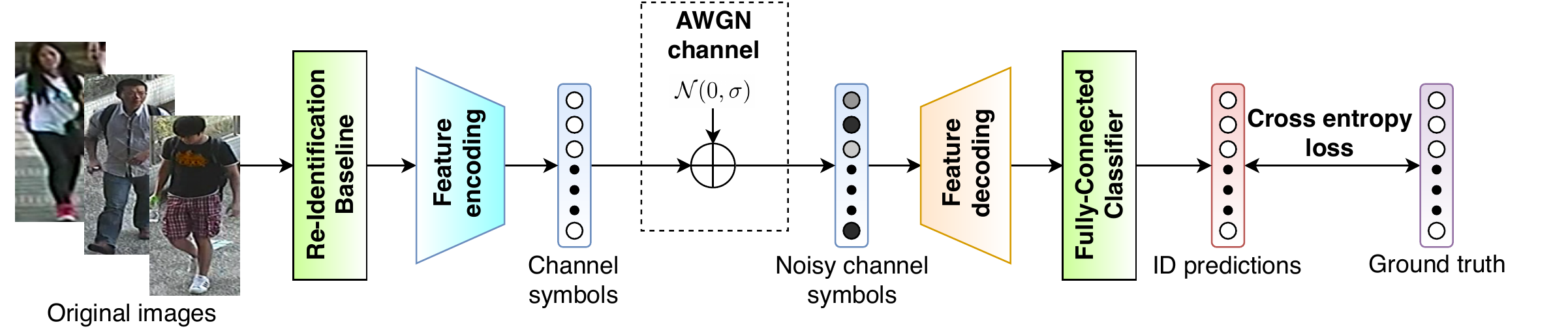}
\end{center}
   \caption{Training of JSCC for person re-ID. The feature vector is directly mapped to channel inputs. Noisy received signal is decoded and processed by a fully-connected layer to obtain ID predictions, which are then compared to the ground truth.}
\label{fig:approach1}
\end{figure}

\subsection{JSCC of Feature Vectors}
\label{subs:approach1}

\begin{figure}[]
    \centering
    \begin{subfigure}[b]{0.47\linewidth}
        \centering
        \includegraphics[width=\textwidth]{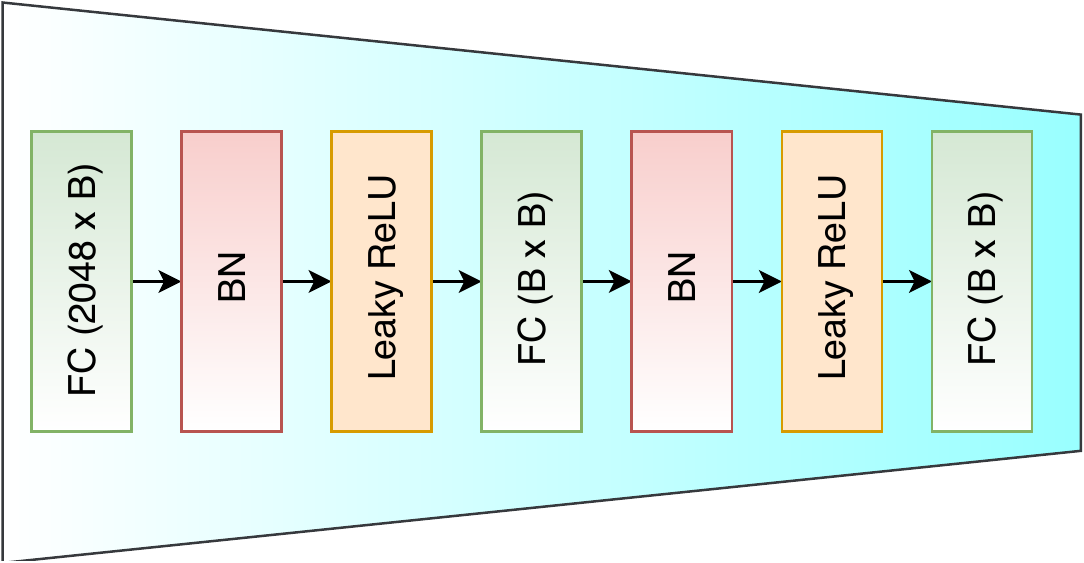}
        \caption{Encoder}
        \label{fig:encoder}
    \end{subfigure}
    \hspace{10pt}
    \begin{subfigure}[b]{0.47\linewidth}
        \centering
        \includegraphics[width=\textwidth]{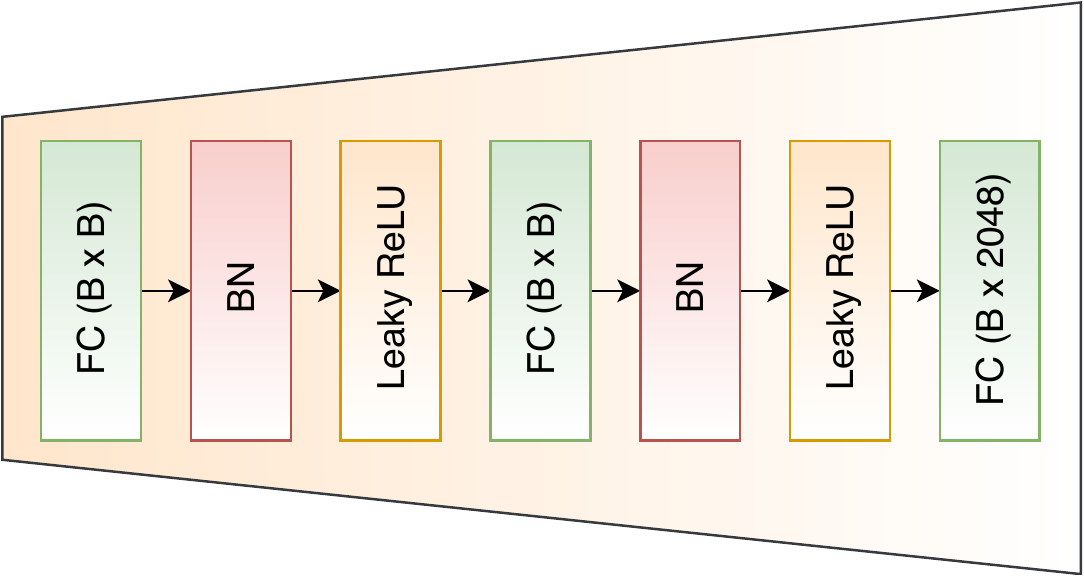}
        \caption{Decoder}
        \label{fig:decoder}
    \end{subfigure}
   
    \caption{Proposed encoder and decoder architecture for the JSCC scheme. In the encoder, dimensionality reduction is performed by the first fully-connected layer, which is inverted at the decoder.}
\end{figure}

In the proposed JSCC approach, called JSCC AE and illustrated in  Fig. \ref{fig:approach1},  we use the re-ID baseline to produce the feature vector for a given input. The feature vector is mapped directly to the channel input symbols via a multi-layer fully-connected encoder (Fig. \ref{fig:encoder}). We set the dimensionality of the channel input vector to $B$, which denotes the available channel bandwidth. We will consider small $B$ values modeling stringent bandwidth and latency constraints, typical for surveillance applications. This low-dimensional representation is normalized to satisfy the average power constraint of $P=1$, and transmitted over an AWGN channel with different SNR values. The noisy channel output vector at the receiver is mapped back to the high-dimensional feature space by decoder (Fig. \ref{fig:decoder}), which mirrors the architecture of the encoder. The resulting feature vector is compared to the database feature vectors to find the nearest neighbour.

Our training strategy consists of three steps. First, we attach a single fully-connected layer at the end of the re-ID baseline that maps $2048$-dimensional feature vectors directly to the class predictions. We then pre-train the network for 30 epochs with batch size of 16, using cross-entropy between class predictions and the ground truth as the loss function. In the second step we use the pretrained re-ID baseline to extract features from all the images in the training dataset. We use these features as inputs to the proposed autoencoder network. We train the autoencoder using $L_1$-loss between the feature vectors and vectors reconstructed by the decoder. We train the encoder and the decoder for 200 epochs with SGD optimizer, learning rate $0.1$, reduced to $0.01$ after 150 epochs, and momentum of $0.9$. We further apply $L_2$ regularizer to the autoencoder model, weighted by $5\cdot 10^{-4}$. Finally, we train the whole network jointly, the autoencoder and the re-ID baseline, for 30 epochs, using the cross-entropy loss with learning rate $0.01$, and for further 10 epochs with learning rate of $0.001$, applying the same optimizer and $L_2$ regularization as in the previous two steps.
Note that we use the fully-connected layer removed from the re-ID baseline after the first step to map feature vectors to class predictions.

We also investigate replacing the feature encoder and decoder in the JSCC architecture in Fig \ref{fig:approach1} by a single fully-connected layer and an identity mapping, respectively. We call the revised model JSCC FC. We train the whole network end-to-end for $50$ epochs with cross-entropy loss, learning rate of $0.01$, reduced to $0.001$ after 30 epochs, and a momentum of $0.9$. We also apply $L_2$ regularization, weighted by $5\cdot 10^{-4}$, to all the parameters, including ResNet-50, feature encoder and fully-connected classifier.

\section{Results}

In this section we will evaluate the performance of the proposed JSCC AE and JSCC FC architectures, and compare with that of the digital scheme presented in Section \ref{subs:approach3}. Before presenting the results, we will first discuss the experimental setup and  the dataset used for the evaluations.

\subsection{Experimental setup}
In order to measure the performance of the re-ID task, we employ widely used CUHK03 \cite{cuhk03} benchmark for person re-ID that contains $14096$ images of $1467$ identities taken from two different camera views. The evaluation measure is the top-1 recognition accuracy, which calculates the fraction of correct IDs at the top of the ranked list retrieved for each query.

\begin{figure}[]
    \centering
    \begin{subfigure}[t]{0.49\textwidth}
        \centering
        \includegraphics[width=\textwidth]{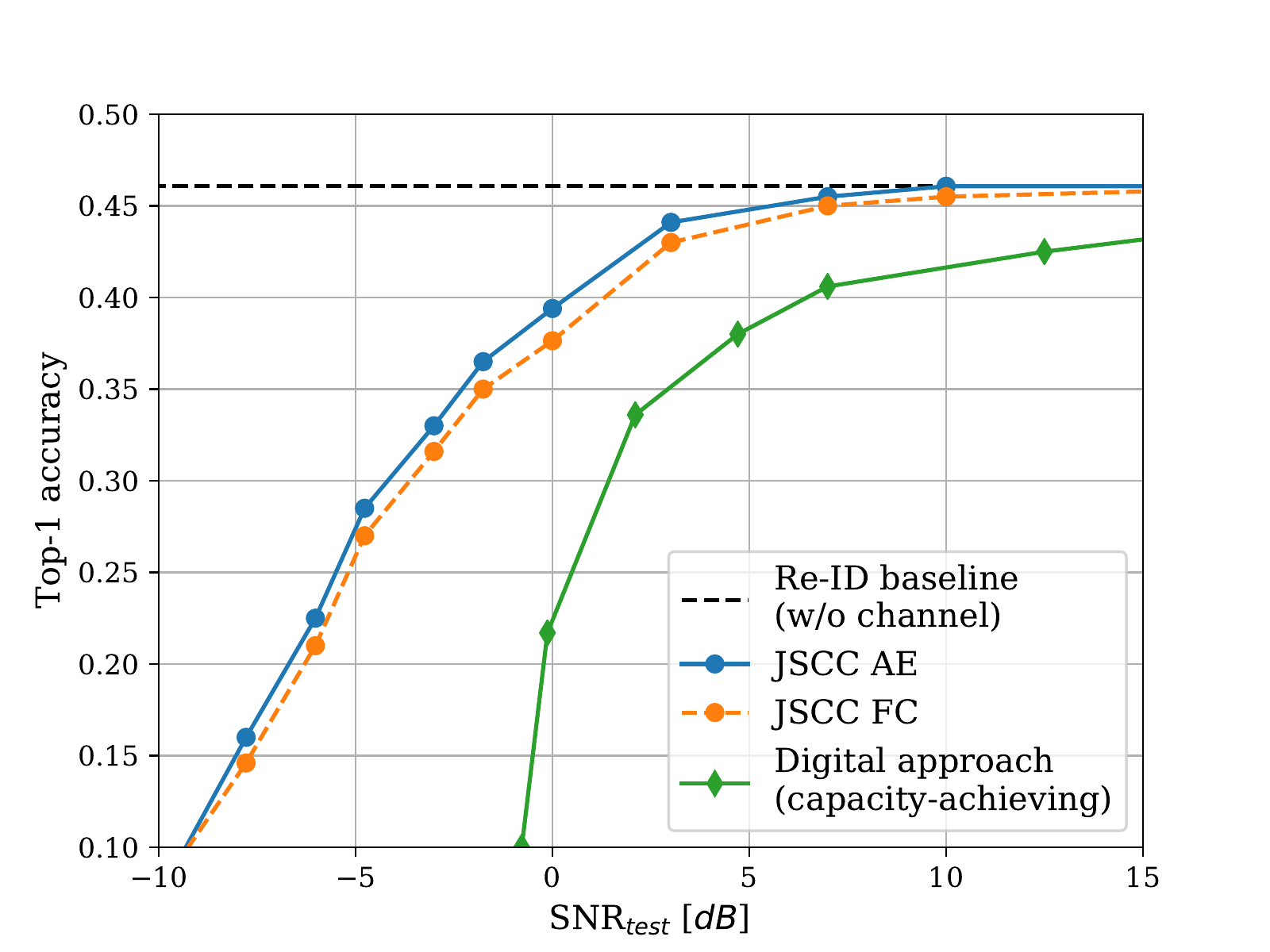}
        \caption{Different approaches}
        \label{fig:results_noiseless_a}
    \end{subfigure}
    \begin{subfigure}[t]{0.49\textwidth}
        \centering
        \includegraphics[width=\textwidth]{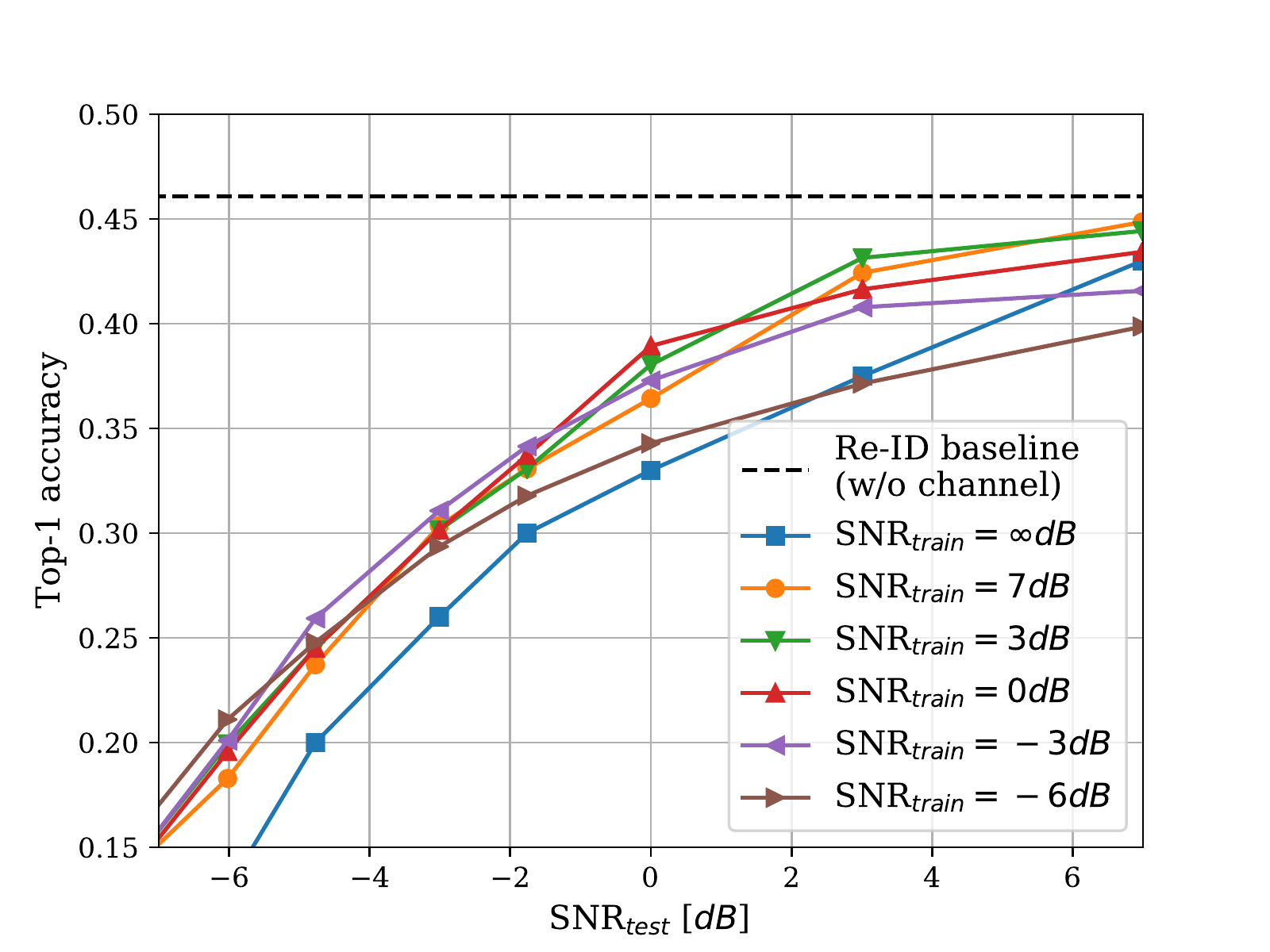}
        \caption{Different training SNRs}
        \label{fig:results_noiseless_b}
    \end{subfigure}
    \caption{Person re-ID top-1 accuracy  as a function of channel SNR when the training and test SNR are the same in (a), and when they differ (b).}
    \label{fig:results_noiseless}
\end{figure}

For the JSCC AE and JSCC FC schemes we consider different channel SNRs for training, varying between $\text{SNR}_{train} = -10\mathrm{dB}$ and $\text{SNR}_{train} = \infty \mathrm{dB}$, which corresponds to zero noise power. In the digital scheme, we allow for different dimensionality of the latent representation, between $64$ and $512$, estimate and minimize its entropy in the training phase by varying the value of parameter $\lambda$. In the testing phase we perform rounding to the nearest integer on each element of the latent representation and arithmetic coding, which is based on the probabilistic model learned by the entropy estimator, as described in \cite{balle2018variational}. This model assigns a probability estimate to each quantized symbol, which is then passed to the arithmetic encoder. We further calculate the average number of bits required to encode the latent representations, and evaluate the corresponding SNR to deliver that many bits to the receiver, assuming capacity-achieving codes (which is a loose bound on the real performance as practical codes are far from the capacity bound in the short blocklength regime).
\subsection{Performance for different methods}

We plot the accuracy achieved by various schemes as a function of the test SNR in  Fig.~\ref{fig:results_noiseless_a}. It is clear that JSCC provides a significant gain compared to the digital approach (despite assuming capacity-achieving channel codes), and it meets the noiseless bound at sufficiently high SNR values. Among the two JSCC architectures, the autoencoder-based approach outperforms the fully-connected encoder without a decoder. The low accuracy of the scheme without decoding may stem from the fact that the noise directly affects the low-dimensional feature vector, while the autoencoder-based scheme introduces certain level of denoising, which improves the feature estimates at the receiver.

In addition to its poor performance, digital transmission also suffers from the \textit{cliff effect}; that is, its performance degrades sharply if the experienced channel SNR falls below the target. On the other hand, the JSCC approach can better accommodate channel variations. In Fig. \ref{fig:results_noiseless_b}, we plot the accuracy results as a function of the test SNR, for networks trained for different $\mathrm{SNR_{train}}$ values. We can see that the JSCC scheme achieves graceful degradation with channel quality. This also means that there is no need to train a separate network for every SNR value. The networks trained at a single moderate SNR value of $0 \mathrm{dB}$ or $-3 \mathrm{dB}$ give satisfactory results tested for a wide range of SNR values. For the considered scenario, channel SNR of at least $10 \mathrm{dB}$ is needed to recover the original re-ID baseline accuracy.

\subsection{Performance for different bandwidths}

\begin{figure}[]
    \centering
    \includegraphics[width=0.8\textwidth]{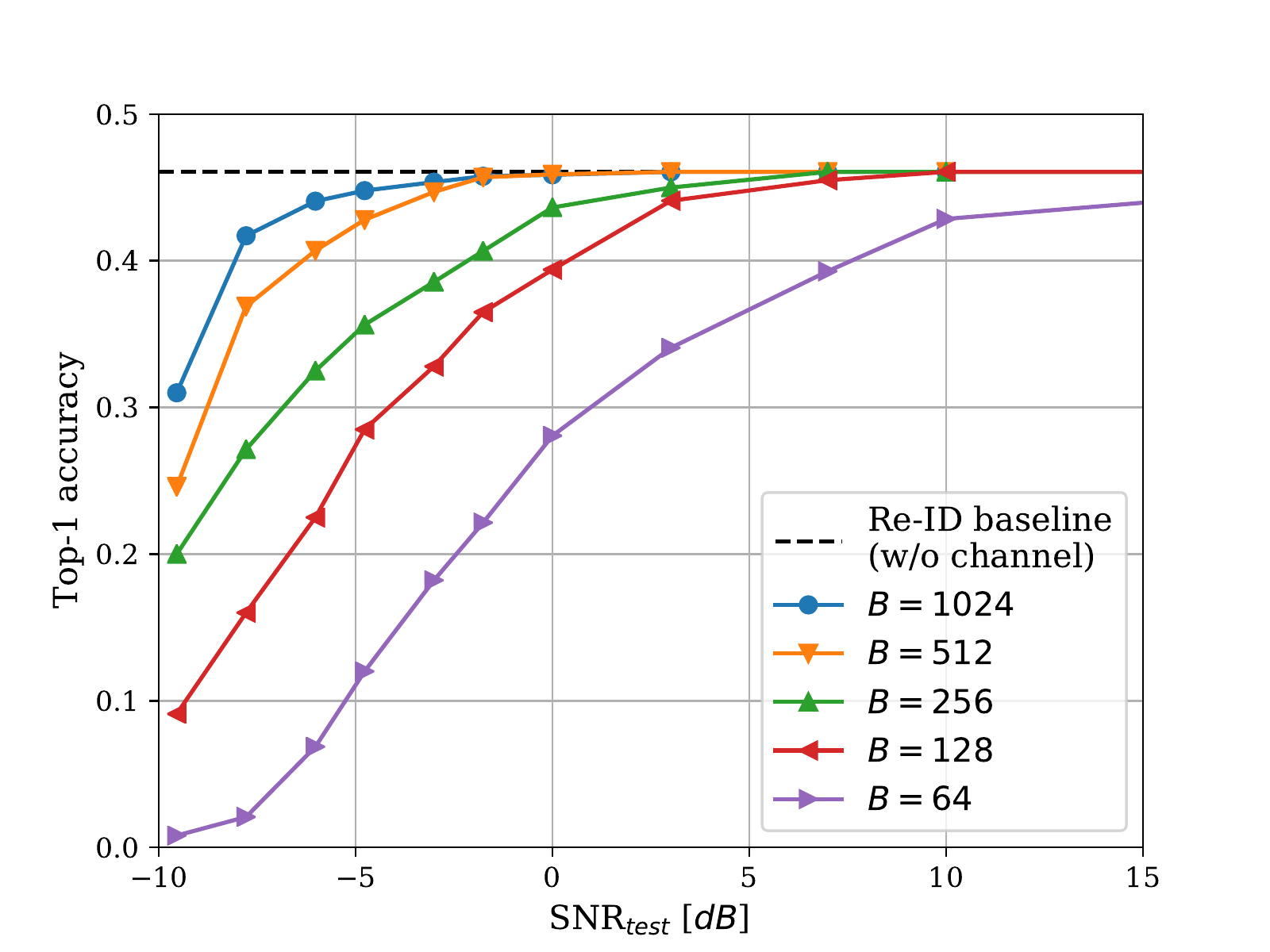}
    \caption{Person re-ID top-1 accuracy  as a function of channel SNR when the training and test SNR are the same for different bandwidths.}
    \label{fig:results_bandwidth}
\end{figure}

In the last experiment we investigate the effect of the channel bandwidth $B$ on the re-ID performance. The accuracy as a function of channel SNR is plotted in Fig. \ref{fig:results_bandwidth} for different channel bandwidth values of $64, 128, 256, 512$ and $1024$. It can be seen that the accuracy and robustness increases significantly with the bandwidth, but the relative gain becomes smaller as we approach the original feature vector dimension. Therefore, from accuracy-bandwidth trade-off perspective the best choice is to aim for a bandwidth of $B=512$ or $B=256$ as they provide a  significant accuracy gain, while still operating over a reasonable bandwidth.

\section{Conclusions}

In this work we studied image retrieval over wireless channels. We We first introduced a digital approach using a state-of-the-art deep image compression algorithm adapted to our problem, followed by capacity-achieving channel codes. We then proposed a JSCC scheme for robust transmission of feature vectors over an extremely limited channel bandwidth. We show that the proposed autoencoder-based JSCC scheme achieves superior results in comparison to the digital scheme and the JSCC scheme trained without the feature decoding phase. This result shows that DNN-based JSCC schemes will be essential to meet the harsh latency and bandwidth constraints of retrieval applications over wireless links. 

\clearpage

\bibliographystyle{IEEEbib}
\bibliography{refs}

\end{document}